\documentclass[12pt]{iopart}
\pdfoutput=1
\usepackage{iopams}
\usepackage{cite}
\usepackage{graphicx}

\newcommand{\ket}[1]{|{#1}\rangle}
\newcommand{\bra}[1]{\langle{#1}|}

\newcommand{\ex}{\bi{e}_x}                            
\newcommand{\ey}{\bi{e}_y}                            
\newcommand{\ez}{\bi{e}_z}                            
\def\Rb87{^{87}\mathrm{Rb}}                           

\begin{document}

\title{Position-dependent spin-orbit coupling for ultracold atoms}

\author{S-W~Su$^{1}$, S-C~Gou$^{1,2}$, I-K~Liu$^{1}$, I~B~Spielman$^{3,4}$,
L~Santos$^{5}$, A~Acus$^{6}$, A~Mekys$^{6}$, J~Ruseckas$^{6}$
and G~Juzeli\={u}nas$^{6}$}

\address{$^{1}$ Department of Physics and Graduate Institute of Photonics,
National Changhua University of Education, Changhua 50058, Taiwan}

\address{$^{2}$ Physics Division, National Center for Theoretical Sciences,
Hsinchu 30013, Taiwan}

\address{$^{3}$ Joint Quantum Institute, University of Maryland, College Park,
Maryland 20742-4111, 20742, USA}

\address{$^{4} $National Institute of Standards and Technology, Gaithersburg,
Maryland 20899, USA}

\address{$^{5}$ Institut f\"ur Theoretiche Physik, Leibniz Universit\"at Hannover,
Appelstr. 2, DE-30167 Hannover, Germany}

\address{$^{6}$ Institute of Theoretical Physics and Astronomy, Vilnius University,
A. Go\v{s}tauto 12, Vilnius LT-01108, Lithuania}

\eads{\mailto{Gediminas.Juzeliunas@tfai.vu.lt}, \mailto{ian.spielman@nist.gov}}

\begin{abstract}
We theoretically explore atomic Bose-Einstein condensates (BECs) subject
to position-dependent spin-orbit coupling (SOC). This SOC can be produced
by cyclically laser coupling four internal atomic ground (or metastable)
states in an environment where the detuning from
resonance depends on position. The resulting spin-orbit coupled BEC
phase-separates into domains, each of which contain density
modulations---stripes---aligned either along the $x$ or $y$ direction. In
each domain, the stripe orientation is determined by the sign of the
local detuning. When these stripes have mismatched spatial periods along domain boundaries, non-trivial topological spin textures form at the interface, 
including skyrmions-like spin vortices and anti-vortices.
In contrast to vortices present in conventional rotating BECs,
these spin-vortices are stable topological defects that are not present in the corresponding homogenous stripe-phase spin-orbit coupled BECs. 
\end{abstract}
\submitto{\NJP}
\maketitle

\section{Introduction}

A number of novel schemes have been proposed to create spin-orbit coupling
(SOC) of Rashba-Dresselhaus type for ultracold atoms by illuminating them with
laser fields
\cite{Ruseckas2005,Stanescu2007,Jacob2007,Juzeliunas2008PRA,Campbell2011,Dalibard2011,Anderson2012PRL,Zhai2012JMPB,Galitski2013,LiuLawNg,Zhai2014-review,Goldman2014}
or by applying pulsed magnetic field gradients \cite{Anderson2013,Xu2013}. SOC
significantly enriches the system, for example leading to non-conventional
Bose-Einstein condensates (BECs)
\cite{Stanescu2008,Wu2011halfvortex,Sinha2011,Wang2010,Zhai2012JMPB,Zhai2014-review}
or Fermi gases with altered pairing~
\cite{Vyasanakere:2011,Vyasanakere:2011bis,Yu:2011,Jiang2011}. Here we extend
current studies by focusing on pseudospin-1/2 BECs subject to spatially
inhomogeneous SOC, and show that these systems form strip-domains interrupted by
non-trivial topological structures at the domain boundaries.

In this article, we focus on real atomic systems from which we simultaneously
identify a pseudospin-1/2 system, and induce SOC with the desired
spatial dependence. This must be achieved using terms naturally entering
into the bare atomic Hamiltonian. Here we show that this may be realized by first creating SOC by cyclically coupling
together four ground (or metastable) atomic states via two-photon
Raman transitions, and then by spatially varying the detuning from
two-photon Raman resonance. We present an explicit construction for
$\Rb87$ in which SOC and the desired spatial dependance coexist.
We then explore the resulting equilibrated pseudospin-1/2 spin-orbit coupled
Bose-Einstein condensates (SOBECs) resulting from this construction.
These SOBECs contain domains of differently oriented stripe
phases. When the stripe's projection onto the domain-boundaries are
spatially mismatched (see Fig.~\ref{fig:PhysicalPicture}), arrays
of non-trivial topological structures such as vortices and anti-vortices in the spin degree of freedom -- skyrmions -- form.

The paper is organized as follows: in Sec.~\ref{sec:physical}
we present a simple physical picture elucidating implications of the
position-dependent SOC; in Sec.~\ref{sec:SOC} we formulate the light-atom
interaction for the specific example of $\Rb87$, and derive the associated
position-dependent spin-orbit coupled Hamiltonian for ground-state
atoms; and in Sec.~\ref{sec:BEC} we use the Gross-Pitaevskii equation
(GPE) to study the ground state structure of these inhomogeneous
systems. Finally, Sec.~\ref{sec:conclusion} summaries our findings.

\section{Physical picture}

\label{sec:physical}

\begin{figure}
\begin{center}
\includegraphics[width=5in]{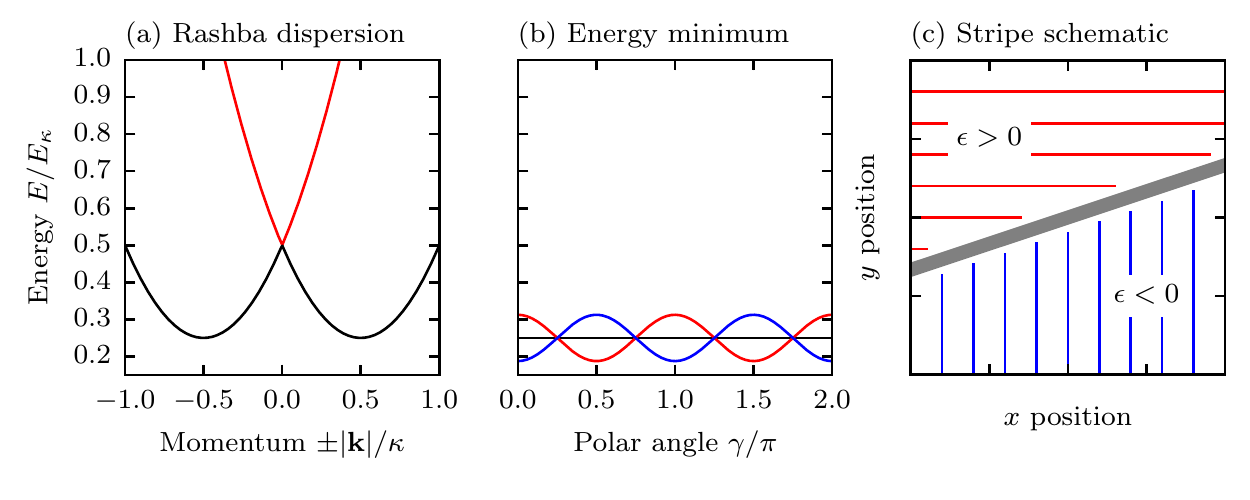}
\end{center}
\caption{Spin-orbit coupled dispersion relations and spatial stripe patterns.
(a) Pure Rashba dispersion plotted along the radial direction $|\bi{k}|/\kappa$.
(b) Energy at the radial minima of the SOC dispersion (i.e.,
$|\bi{k}|/\kappa\approx1/2$) plotted as a function of the polar angle $\gamma$ for
$\epsilon=1/8$ (red), $\epsilon=0$ (black), and $\epsilon=-1/8$ (blue). (c)
Representative stripe pattern showing mismatched stripe periods as projected
onto the domain boundary, resulting from equal period stripes aligned along
$\ex$ and $\ey$.}
\label{fig:PhysicalPicture} 
\end{figure}

Before delving into a detailed discussion of specific atomic systems,
we first discuss the qualitative physics leading to the formation
of topological defects in our system. Our focus is on spin-$1/2$
SOBECs containing mostly Rashba-type SOC contaminated by a small tunable
contribution of Dresselhaus-type SOC; together, these are parametrized
by a non-Abelian vector potential $\bi{A}$, and are described
by the single particle Hamiltonian 
\[
H =\frac{1}{2m}\left(\hbar\bi{k}-\bi{A}\right)^{2}\,,\quad\mathrm{where}\quad
\bi{A}=\frac{\hbar\kappa}{2}\left[\left(1-\epsilon\right)\sigma^{x}\ex
-\left(1+\epsilon\right)\sigma^{y}\ey\right],
\]
$\sigma^{x,y,z}$ are the Pauli operators, and $I$ is the identity. Here, $m$ is
the atomic mass; $\hbar\bi{k}$ is the momentum; $\kappa\geq0$ both
describes the Rashba SOC strength and defines the energy
$E_{\kappa}=\hbar^{2}\kappa^{2}/2m$; and lastly, $\epsilon\kappa$ describes the
Dresselhaus SOC strength. The eigenvalues of this Hamiltonian (shown in
Fig.~\ref{fig:PhysicalPicture}a) are
\begin{eqnarray}
\fl E_{\pm}(\bi{k}) =E_{\kappa}\left\{ \left[\left(\frac{\bi{k}}{\kappa}\right)^{2}
+\frac{1}{2}\left(1+\epsilon^{2}\right)\right]
\pm\left[\left(\frac{\bi{k}}{\kappa}\right)^{2}\left(1+\epsilon^{2}\right)
-2\left(\frac{k_{x}}{\kappa}\right)^{2}\epsilon
+2\left(\frac{k_{y}}{\kappa}\right)^{2}\epsilon\right]^{1/2}\right\} \nonumber \\
\approx E_{\kappa}\left\{ \left[\left(\frac{\bi{k}}{\kappa}\right)^{2}
+\frac{1}{2}\right]\pm\left|\frac{\bi{k}}{\kappa}\right|
\left[1-\epsilon\frac{k_{x}^{2}-k_{y}^{2}}{\bi{k}^{2}}\right]\right\} +O(\epsilon^{2}),
\end{eqnarray}
where the second equation is valid to linear order in $\epsilon$. For
$\epsilon=0$, these energies depend only on $\left|\bi{k}\right|$, so the
ground state (minimum of lower energy band, $E_{-}(\bi{k})$) is macroscopically
degenerate on the ring $|\bi{k}/\kappa|=1/2$. Figure~\ref{fig:PhysicalPicture}b
plots the energy minimum of radial cuts through SOC dispersion relations as a
function of the polar angle $\gamma$, where $\bi{k}=|\bi{k}|
\left(\cos\gamma\ex+\sin\gamma\ey\right)$; the black line, independent of
$\gamma$, indicates the degenerate ground states of the Rashba Hamiltonian. This
massive degeneracy is lifted when $\epsilon\neq0$. In this case, the dispersion
is two-fold degenerate with minima at $\bi{k}_{\pm}/\kappa=\mp(1+\epsilon)\ey/2$
for $\epsilon>0$ (red curve in
Fig.~\ref{fig:PhysicalPicture}b) and $\bi{k}_{\pm}/\kappa=\pm(1-\epsilon)\ex/2$
for $\epsilon<0$ (blue curve in Fig.~\ref{fig:PhysicalPicture}b). The
corresponding minimum energy eigenstates have their pseudospin aligned along $\bi{k}_{\pm}$.

Under many realistic physical conditions, a SOBEC will Bose-condense into both
of these minima simultaneously~\cite{Wang2010,Yu}, and the spatial interference
between these two states, differing in momentum by $\delta k\approx\kappa$, will
generate stripes in the atomic spin density with spatial period $2\pi/\kappa$.
These stripes are aligned parallel to $\ex$ for $\epsilon>0$ and parallel to
$\ey$ for $\epsilon<0$.

Here we study physical systems where the magnitude of the Dresselhaus SOC
$\kappa\epsilon$ varies linearly along a direction in the $\ex$ - $\ey$ plane
defined by the unit vector $\bi{e}=\cos\theta\ex+\sin\theta\ey$. In the
half-plane with $\epsilon>0$ we expect horizontal stripes and in the half-plane
with $\epsilon<0$ we expect vertical stripes (schematically shown in
Fig.~\ref{fig:PhysicalPicture}c), and we ask: how are these different patterns
of stripes linked at the boundary line $0=x\cos\theta+y\sin\theta$ delineating
the two domains (grey line in Fig.~\ref{fig:PhysicalPicture}c). This seemingly
simple question is nontrivial because the horizontal stripes ($\epsilon>0$) have
period $d_{+}=|2\pi/\kappa\sin\theta|$ projected onto the delineating line,
while vertical stripes ($\epsilon<0$) have period
$d_{-}=|2\pi/\kappa\cos\theta|$ along the delineating line (see
Fig.~\ref{fig:PhysicalPicture}c): when $|\cos\theta|\neq|\sin\theta|$ stripes
must terminate or originate at the domain boundary, leading to the formation of
pinned topological defects.

\section{Position-dependent SOC}

\label{sec:SOC}

\subsection{The electronic Hamiltonian and its eigenstates}

\label{sec:Ham}

\begin{figure}
\begin{center}
\includegraphics[width=4.5in]{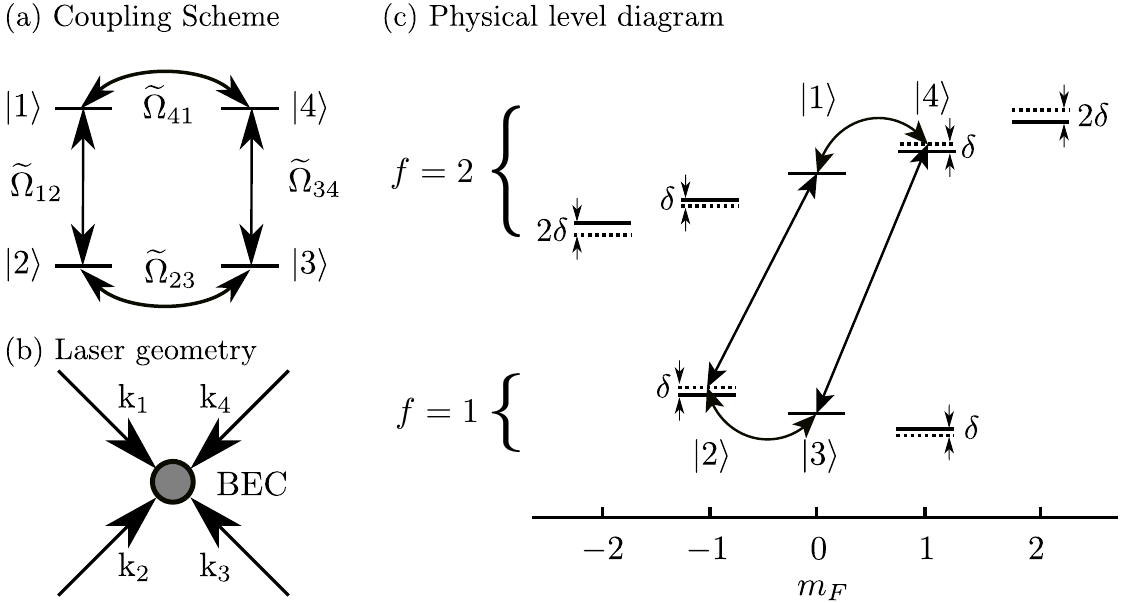}
\end{center}
\caption{Proposed four level coupling scheme. (a) Coupling diagram showing our
four cyclically coupled internal atomic states. (b) Spatial geometry of the
coupling laser beams driving the two-photon Raman transitions. (c) Realization
of the closed loop scheme in $\Rb87$ using Raman transitions between magnetic
sub-levels of the $f=1$ and $f=2$ hyperfine manifolds. Here $\hbar\delta=g_F \mu_B B_0$ is the detuning of the atomic levels $|2\rangle$ and $|4\rangle$ from two-photon Raman resonance due to an inhomogeneous magnetic field. Each line or curve connecting the bare states depicts a two-photon Raman transition.
}
\label{fig:Closed-loop-level} 
\end{figure}

Our inhomogeneous SOC may be created using any atom with four internal
ground (or metastable) states $\{|1\rangle,|2\rangle,|3\rangle,|4\rangle\}$
that can be coupled in the cyclic manner shown in Fig.~\ref{fig:Closed-loop-level}a.
In $\Rb87$ these might be the four ground hyperfine states~\cite{Campbell2011}  illustrated in
Fig.~\ref{fig:Closed-loop-level}c:
$\ket{f=1,m_{F}=0,-1}$ and $\ket{f=2,m_{F}=0,+1}$. In that case the four states
are Raman coupled with the position-dependent couplings
\begin{equation}
\tilde{\Omega}_{j}=\Omega_{j}\exp\left\{i[(\bi{k}_{j+1}
-\bi{k}_{j})\cdot\bi{r}-\phi_{j}]\right\}
\label{Omega_tilde_j}
\end{equation}
with amplitude $\Omega_{j}$, recoil momentum
$\bi{k}_{j+1}-\bi{k}_{j}$ and phase shift $\phi_{j}$. Here
\begin{equation}
\bi{k}_j =\kappa\{\ex \cos(\pi j/2) + \ey \sin(\pi j/2)\}\,,\quad\,j=1,2,3,4,
\label{k_j}
\end{equation}
is the wave-vector of the $j^{\mathrm{th}}$ Raman laser field,
$\kappa$ being its length. This coupling scheme can be
realized using the combination of $\pi$ and $\sigma$ polarized laser fields laid
out in Ref.~\cite{Campbell2011}. The linear Zeeman shift, from a biasing
magnetic field $\bi{B}_{0}=B_{0}\ez$, is rendered position-dependent by virtue
of an additional magnetic field gradient $\bi{B}^{\prime}=
B^{\prime}(\bi{r}\cdot\bi{e})\ez$, linearly varying in the $\ex$ - $\ey$ plane along the direction
$\bi{e}=\cos\theta\ex+\sin\theta\ey$. The combination $\bi{B}_{0}
+\bi{B}^{\prime}$ provides a controllable detuning
$\delta=\delta_{0}+\delta^\prime(\bi{r}\cdot\bi{e})$ from Raman resonance to
the states $|2\rangle$ and $|4\rangle$, see Fig.~\ref{fig:Closed-loop-level}c.
Physically this can be realized by using atomic magnetic levels shown in
Fig.~\ref{fig:Closed-loop-level}c where one pair of states is field insensitive
and the other pair share essentially the same
$E_{\mathrm{Z}}=g_{F}\mu_{\mathrm{B}} |{\bf B}| m_{F}$
Zeeman shift, where $\mu_{\mathrm{B}}$ is the Bohr magneton and $g_{F}$ is the Land{\'e} g-factor (opposite in sign for the $f=1$ and $f=2$ mailfolds).

The scheme of cyclically coupled states, shown in
Fig.~\ref{fig:Closed-loop-level}, is formally equivalent to a four-site lattice
with periodic boundary conditions, i.e., $|5\rangle\equiv|1\rangle$. In terms of
the position-dependent states
$|\tilde{j}\rangle\equiv|\tilde{j}\left(\bi{r}\right)\rangle=\exp\left(-i\bi{k}_{j}\cdot\bi{r}\right)|j\rangle$,
the Hamiltonian describing the internal atomic degrees of freedom is
\begin{equation}
\frac{\hat{H}_{e}}{\hbar} =\tilde\delta I+\sum_{j=1}^{4}\tilde\delta(-1)^{j}|\tilde{j}\rangle
\langle\tilde{j}|-\sum_{j=1}^{4}\left[\Omega_{j}e^{-i\phi_{j}}|\tilde{j}\rangle
\langle\widetilde{j+1}|+\mathrm{H.c.}\right]\,.\label{eq:H_e}
\end{equation}
where the contribution from the atom-light detuning is represented in
terms of the overall shift of the energy zero $\tilde\delta I$ and an alternating
detuning $\tilde\delta(-1)^{j}$, where $\tilde\delta = \delta/2$. This corresponds exactly to the
experimental situation illustrated in Fig.~\ref{fig:Closed-loop-level}c, where
the levels $\ket{2}$ and $\ket{4}$ are shifted by $2\tilde\delta$, whereas the levels $\ket{1}$
and $\ket{3}$ are unaffected.  Because $\tilde\delta$ depends on
position, the energy offset $\tilde\delta$ cannot be removed by globally shifting the zero of energy, as in Ref.~\cite{Campbell2011}. Instead the local energy shift from  $\tilde\delta(\bi{r})$ is implicitly incorporated into the trapping potential $V(r)$ featured in Eq.~(\ref{energy}): our linearly varying detuning simply shifts the location of the harmonic potential's minimum.

The phases of the laser fields are taken such that $\sum_{j}\phi_{j}=\pi$, implying
that an atom acquires a $\pi$ phase shift upon traversing the closed-loop
$|1\rangle\rightarrow|2\rangle\rightarrow|3\rangle\rightarrow|4\rangle\rightarrow|1\rangle$
in state space. For zero detuning ($\tilde\delta=0$) and equal Rabi frequencies
($\Omega_{j}=\Omega$) the eigenfunctions and corresponding eigenvalues are
\begin{equation}
\fl\ket{\chi_{q}} =\frac{1}{2}\sum_{p=1}^{4}\rme^{i\pi qp/2}|\tilde{p}\rangle\,,\quad
\mathrm{and}\quad\varepsilon_{q} =-2\Omega\cos\left[\frac{\pi}{2}\left(q-\frac{1}{2}
\right)\right]\,,\quad\mathrm{with}\quad q\in\left\{0,\ldots,3\right\}\,.
\label{eq:Eigen-functions-delta_0}
\end{equation}
In the $\left\{ \ket{\chi_{q}}\right\} $ basis, the internal Hamiltonian
\begin{equation}
\frac{\hat{H}_{e}}{\hbar} =\sum_{q=0}^{3}\varepsilon_{q}|\chi_{q}\rangle\langle\chi_{q}|
+\tilde\delta\sum_{q=0}^{3}|\chi_{q}\rangle\langle\chi_{q-2}|
\label{eq:ham-eigen-basis}
\end{equation}
has two pairs of degenerate eigenstates
\begin{equation}
|\downarrow,\pm\rangle =a_{\mp}|\chi_{0}\rangle\pm a_{\pm}|\chi_{2}\rangle\,,\quad
\mathrm{and}\quad |\uparrow,\pm\rangle =
a_{\mp}|\chi_{1}\rangle\pm a_{\pm}|\chi_{3}\rangle\,,
\label{eq:Dressed_states_lower}
\end{equation}
labeled by the pseudospin index $\uparrow,\downarrow$ and by their energies
$\pm\hbar\sqrt{\tilde\delta^{2}+2\Omega^{2}}$, where
\begin{equation}
a_{\pm} =\sqrt{\frac{1}{2}\pm\frac{\sqrt{2}\Omega}{2\sqrt{\tilde\delta^{2}+2\Omega^{2}}}}\,.
\label{eq:coef}
\end{equation}

\subsection{Adiabatic motion and spin-orbit coupling}

We are interested in the situation where the separation energy
$2\hbar\sqrt{\tilde\delta^{2}+2\Omega^{2}}$ between the pairs of dressed states
greatly exceeds the kinetic energy of the atomic motion. In that case the atoms
adiabatically move about within each two-fold degenerate manifold of pseudospin
states.   Such adiabatic motion is affected by the matrix-valued geometric vector and
scalar potentials $\bi{A}^{(\pm)}$ and $\Phi^{(\pm)}$ which result from the
position-dependence of the atomic internal dressed states
\cite{Dalibard2011,Goldman2014}.  Here the $\pm$ signs denote to the ground or
excited adiabatic manifold.  The matrix elements of the gauge potentials are
\begin{equation}
\fl\bi{A}^{(\pm)}_{s,s^{\prime}}= i\hbar \bra{s,\pm} \bnabla
\ket{s^{\prime},\pm}\,,\quad\mathrm{and}\quad\Phi^{(\pm)}_{s,s^{\prime}}
=-\frac{\hbar^2}{2m} \sum_{s^{\prime \prime}=\uparrow,\downarrow} \bra{s,\pm}
\bnabla\ket{s^{\prime \prime},\mp} 
\bra{s^{\prime \prime},\mp} \bnabla\ket{s^{\prime},\pm} \,,
\end{equation}
where $s,s^{\prime}\in\left\{\uparrow,\downarrow\right\}$. Using
Eq.~(\ref{eq:Dressed_states_lower}) for the dressed states
$\left|\uparrow,\downarrow;-\right\rangle$, one arrives at the explicit result
for the gauge potentials in the ground-state manifold
\begin{eqnarray}
\bi{A}^{(-)}=\frac{\hbar\kappa}{2}\left[\left(1-\epsilon\right)\sigma^{x}\ex
-\left(1+\epsilon\right)\sigma^{y}\ey\right]\,,
\label{eq:A-detuning}\\
{
\Phi^{(-)} =\frac{\hbar^2}{4m}\eta
\left[\kappa^2 + \eta\frac{(\bnabla\tilde\delta)^2}{4\Omega^2}\right]I\,,
\label{eq:Phi-detuning-full}
}
\end{eqnarray}
with
\begin{equation}
\epsilon=2a_{+}a_{-}=\frac{\tilde\delta}{\sqrt{\tilde\delta^{2}+2\Omega^{2}}}\,,\quad\mathrm{and}\quad
\eta=\left(a_{+}^{2}-a_{-}^{2}\right)^{2}=\frac{1}{1+\tilde\delta^{2}/2\Omega^{2}}\,.
\label{eq:ab__a^2-b^2}
\end{equation}
Since the detuning $\tilde\delta$ varies slowly over the optical wavelength,
the spatial derivatives of the detuning can be neglected in
Eq.~(\ref{eq:Phi-detuning-full}), giving
\begin{equation}
\Phi^{(-)}=\eta\frac{\hbar^2\kappa^2}{4m}I\,.
\label{eq:Phi-detuning}
\end{equation}

When the detuning is much smaller than the Rabi frequency, $\tilde\delta\ll\Omega$,
the lowest order in $\tilde\delta$ contribution to the gauge potentials
$\bi{A}\equiv\bi{A}^{(-)}$ and $\Phi\equiv\Phi^{(-)}$ are linear and
quadratic respectively,
\begin{eqnarray}
\bi{A} \approx \frac{\hbar\kappa}{2}\left[\left(1-\frac{1}{\sqrt{2}}
\frac{\tilde\delta}{\Omega}\right)\sigma^{x}\bi{e}_{x}-\left(1+\frac{1}{\sqrt{2}}
\frac{\tilde\delta}{\Omega}\right)\sigma^{y}\bi{e}_{y}\right]\,,
\label{eq:A-detuning-small}\\
\Phi \approx \frac{\hbar^{2}\kappa^{2}}{4m}
\left(1-\frac{\tilde\delta^{2}}{2\Omega^{2}}\right)I\,.
\label{eq:Phi-detuning-small}
\end{eqnarray}
The effective scalar potential $\Phi$, resulting from the adiabatic elimination
of the excited states, is proportional to the unit matrix and hence provides
only an additional state-independent trapping potential.

The matrix-valued vector potential can be equivalently understood as SOC with
spatially-dependence appearing via the position dependent detuning
$\tilde\delta\equiv\tilde\delta\left(\bi{r}\right)$. For zero detuning, the vector
potential is proportional to $\sigma^{x}\ex-\sigma^{y}\ey$, so the SOC is
cylindrically symmetric. For non-zero detuning the cylindrical symmetry is lost,
leading to the formation of the stripe phases in the SOC BEC along $\ex$ or
$\ey$ as was discussed in Sec.~\ref{sec:physical}.

\section{BEC with position-dependent SOC}

\label{sec:BEC}

\subsection{Equations of motion}

Having now shown how to create inhomogeneous SOC, we shift our focus
to its effects on ground state properties of BECs. At zero temperature,
the mean-field energy functional of a spin-1/2 BEC with SOC is 
\begin{equation}
\fl E\left[\bPsi^{*},\bPsi\right] =\int \rmd\bi{r}\bigg[\Psi^{*}
\frac{\left(\bi{p}-\bi{A}\right)^{2}}{2m}\Psi+V\left(r\right)\left|
\Psi\right|^{2}+\Phi\left(\bi{r}\right)\left|\Psi\right|^{2}+\frac{g}{2}\left|
\Psi\right|^{4}-\mu\left|\Psi\right|^{2}\bigg]\,,\label{energy}
\end{equation}
where
$\bPsi=\left(\psi_{\downarrow},\psi_{\uparrow}\right)^{\mathit{T}}$ is
the spinor (vectorial) order parameter, $V\left(r\right)=m\omega^{2}r^{2}/2$ is the
trapping potential and $g$ is the nonlinear interaction strength. The synthetic
vector and scalar gauge potentials $\bi{A}$ and $\Phi$
[Eqs.~(\ref{eq:A-detuning}), (\ref{eq:Phi-detuning})] depend on the linearly
varying detuning
\begin{equation}
\tilde\delta =\tilde\delta^\prime\left(x\cos\theta+y\sin\theta\right)\label{eq:delta/Omega}
\end{equation}
introduced in Sec.~\ref{sec:Ham}. Here we assume that $V(\bi{r})$ embodies all
external potentials including that resulting from the spatially-dependent energy
offset in Eq.~(\ref{eq:H_e}).

The spinor time-dependent GPE (TDGPE) can be derived via the Hartree variational
principle $i\hbar\partial_{t}\psi_{j}=\delta E/\delta\psi_{j}^{*}$ giving 
\begin{eqnarray}
i\hbar\partial_{t}\psi_{s} & =\sum_{s^{\prime}}
\bigg\{\left[-\frac{\hbar^{2}\nabla^{2}}{2m}+V+\Phi+g\rho
+\frac{\left|\bi{A}\right|^{2}}{2m}-\mu\right]I_{s,s^{\prime}}\nonumber \\
& +i\frac{\hbar^{2}\kappa}{2m}\left[\left(A_{x}\partial_{x}+
\frac{\partial_{x}A_{x}}{2}\right)\sigma_{s,s^{\prime}}^{x}+\left(A_{y}\partial_{y}
+\frac{\partial_{y}A_{y}}{2}\right)\sigma_{s,s^{\prime}}^{y}\right]\bigg\}
\psi_{s^{\prime}}\label{eq:equation of motion}
\end{eqnarray}
where $\rho=|\psi_{\downarrow}|^{2}+|\psi_{\uparrow}|^{2}$ is the total density.
Equation~(\ref{eq:equation of motion}) governs the dynamics of the BECs with
position-dependent SOC, at the mean-field level.

\subsection{Ground-state phases of the SOBEC}

In Sec.~\ref{sec:physical}, we discussed the single particle properties
expected in our mixed Rashba-Dresselhaus spin-orbit coupled system
and noted that when $\epsilon\neq0$ the spectrum is two-fold degenerate
at points $\bi{k}_{\pm}/\kappa=\pm(1+\epsilon)\ey/2$ for $\epsilon>0$
and $\bi{k}_{\pm}/\kappa=\pm(1-\epsilon)\ex/2$ for $\epsilon<0$.
When weak repulsive interactions are included, the bosons can condense either in:
(1) a plane-wave phase (PW) in which one of $\bi{k}_{+}$ or $\bi{k}_{-}$
is macroscopically occupied; or in (2) a standing wave phase (SW,
sometimes called a striped phase) in which the bosons condense into
a coherent superposition of $\bi{k}_{+}$ and $\bi{k}_{-}$. We
focus on the case where the inter- and intra- spin interactions are
identical, for which the ground state is in the SW phase~\cite{Wang2010,Yu}.

The detuning $\tilde\delta$ vanishes along the separatrix $x\cos\theta+y\sin\theta=0$
that delineates the regions with $\tilde\delta>0$ and $\tilde\delta<0$.
Since the wave vectors characterizing the two domains have differing projections
onto the line where $\tilde\delta=0$, novel structures can form to heal the otherwise
discontinuous strip patterns at opposite sides of the separatrix. For example, for $\tan\theta=1$ and $2$,
we expect one-to one and two-to-one connections on the interface, respectively.

We determined the ground state of the SOBEC --
minimizing the energy functional in Eq.~(\ref{energy}) -- by propagating
Eq.~(\ref{eq:equation of motion}) with differing degrees of imaginary time~\cite{Choi1998,Penckwitt2002,Tsubota2002,Billam2014}:
replacing $i\partial_{t}$ by $i\exp[i\zeta]\partial_{t}$ in
Eq.~(\ref{eq:equation of motion}), for $\zeta\in[0,\pi/2]$.  While simulations converged to the same solution for any non-negligible $\zeta$, the resulting damped GPE converges much more rapidly for proper choice of  $\zeta$. We
confirmed that obtained the ground state by the absence of any
time-dependance when we evolve in real time.
We considered a 2D $\Rb87$ BEC
confined in a harmonic potential with frequency $\omega/2\pi=100\,\mathrm{Hz}$. For
computational convenience, we adopt the dimensionless units where the frequency
and length are scaled in units of the trap frequency $\omega/2\pi$ and the
oscillator length $\sqrt{\hbar/m\omega}$, respectively. We employ the Fourier
pseudospectral method with $N_{x}=N_{y}=256$ grid points.

\begin{figure}
\begin{center}
\includegraphics{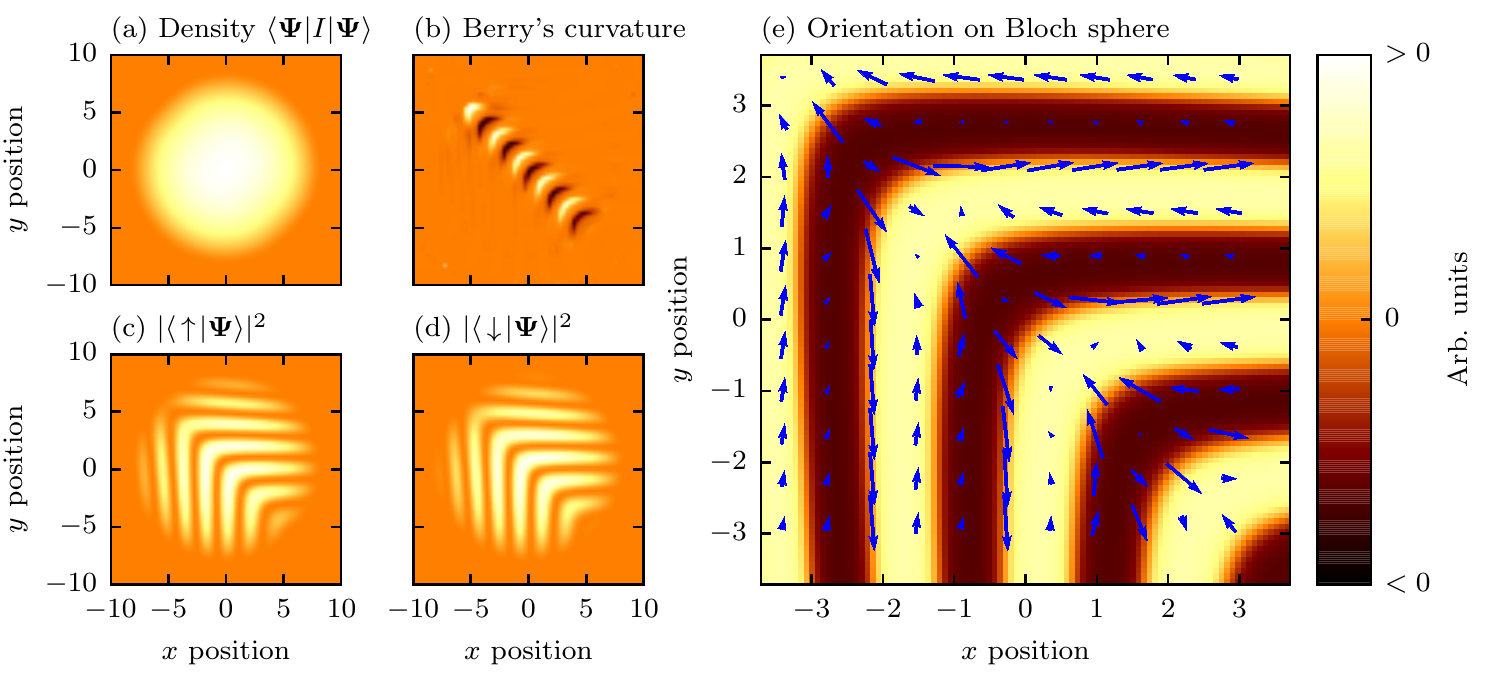}
\end{center}
\caption{The ground state of a spatially dependent SOC BEC with a commensurate
interface. This simulation was performed with $\mu=32\hbar\omega$,
$\tilde\delta^\prime=\Omega/2$, $\kappa=2$, and $\theta=\pi/4$.
(a)--(d) plot the density $\bra{\bPsi}I\ket{\bPsi}$, the Berry's curvature ${\mathcal B}_z$, and the densities in the spin up and spin down components
$|\left<\uparrow\!|\bPsi\right>|^2$, and $|\left<\downarrow\!|\bPsi\right>|^2$.
Each of these quantities varies smoothly across the domain interface,
continuously connecting the the two SW phases.  (e) Orientation of the local spin vector on the Bloch sphere.  The colored background gives the $\ez$ component, while the vector field plots the $\ex$ and $\ey$ components.}
\label{x+y} 
\end{figure}

We first consider the parameters  
$\tilde\delta^\prime=\Omega/2$, $\kappa=2$, $\theta=\pi/4$, and $\mu=32\hbar\omega$ 
which corresponds to $gN_{\uparrow}=gN_{\downarrow}\simeq1500$.  In this case
$\cos\theta=\sin\theta$ and the horizontal and vertical stripes are matched one-to-one
at the boundary. The corresponding ground-state wave function is shown in
Fig.~\ref{x+y}. In this case, the interface lies along the line $x+y=0$ and the
stripes align along $\ey$ for $x+y<0$ and along $\ex$ for $x+y>0$. Clearly, the
stripes in both domains are connected one-to-one across the interface. Moreover,
the orientation of the state along the Bloch sphere smoothly connects the two
phases, as shown in Figs.~\ref{x+y}e. The ground-state structure is
consistent with the prediction of the noninteracting homogeneous system with our
single particle arguments.

These stripes are associated with the local spin vector 
\begin{equation}
{\bf N} = \bra{\bPsi}{\boldsymbol \sigma}\ket{\bPsi}\,,\quad{\rm and\ orientation}\quad{\bf n} = \frac{{\bf N}}{|{\bf N}|}
\end{equation}
precessing on the Bloch sphere (Fig.~\ref{x+y}e) either in the $\ex$-$\ez$ plane (horizontal stripes) or the $\ey$-$\ez$ plane (vertical stripes).  The changing orientation in these precession planes leads to a chain of vortices in the spin degree of freedom: distorted skyrmions and anti-skyrmions.  We quantify the location of these vortices in Fig.~\ref{x+y}b where we plot the local Berry's curvature
\begin{eqnarray}
{\mathcal B}_z &= -\frac{\hbar}{2}\left(\frac{\partial_x n_x \partial_y n_y - \partial_y n_x \partial_x n_y}{n_z}\right)
\end{eqnarray}
which is peaked at the vortex centers.  This clearly shows the ordered chain of skyrmions at the stripe-interface.

\begin{figure}
\begin{center}
\includegraphics{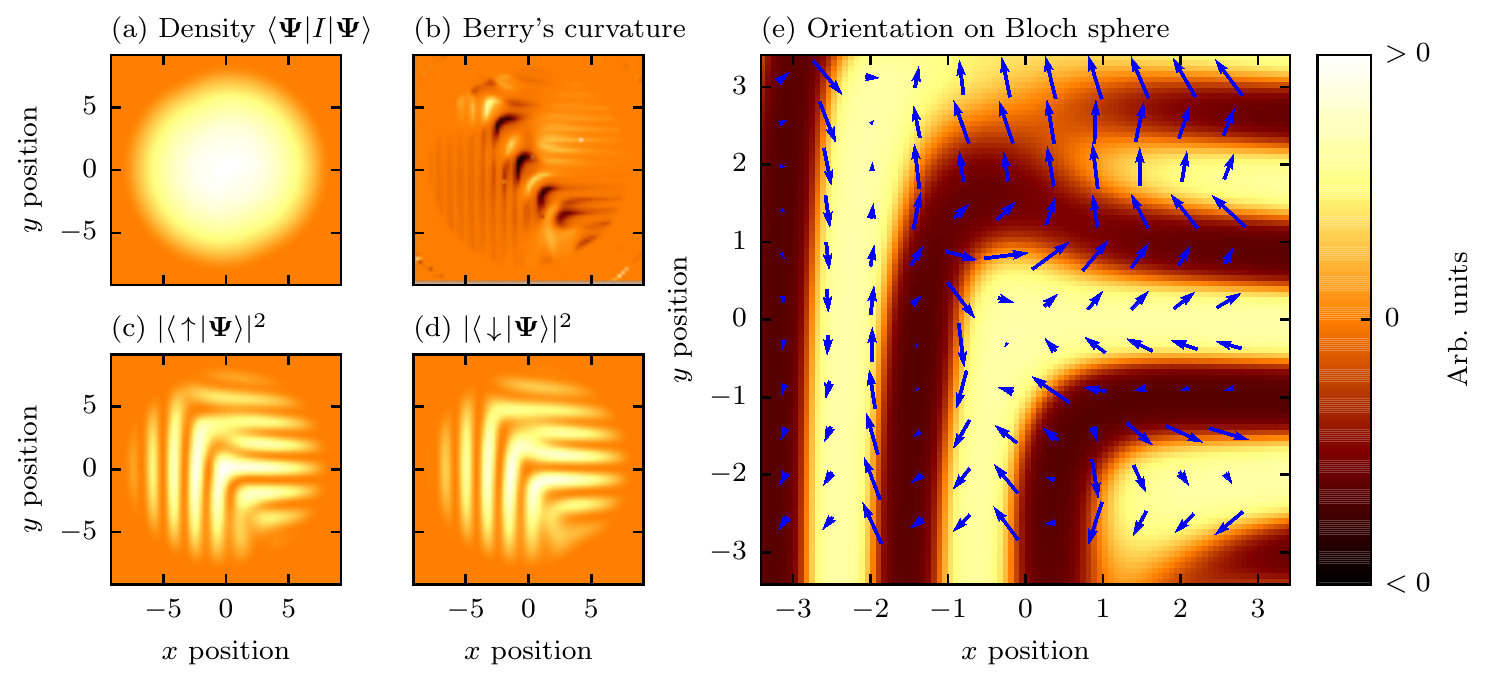}
\end{center}
\caption{The ground state of a spatially dependent SOC BEC with an incommensurate
interface. This simulation was performed with $\mu=32\hbar\omega$,
$\tilde\delta^\prime=\Omega\sqrt{10}/6$, $\kappa=2$ and $\tan\theta=1/2$.
(a)--(d) plot the density $\bra{\bPsi}I\ket{\bPsi}$, the Berry's curvature ${\mathcal B}_z$, and the densities in the spin up and spin down components
$|\left<\uparrow\!|\bPsi\right>|^2$, and $|\left<\downarrow\!|\bPsi\right>|^2$. Because the projection
of the stripe periods along the interface are mismatched, defects
form at the domain interface between the two SW phases.  (e) Orientation of the local spin vector on the Bloch sphere.  The colored background gives the $\ez$ component, while the vector field plots the $\ex$ and $\ey$ components.}
\label{2x+y} 
\end{figure}

\begin{figure}
\begin{center}
\includegraphics{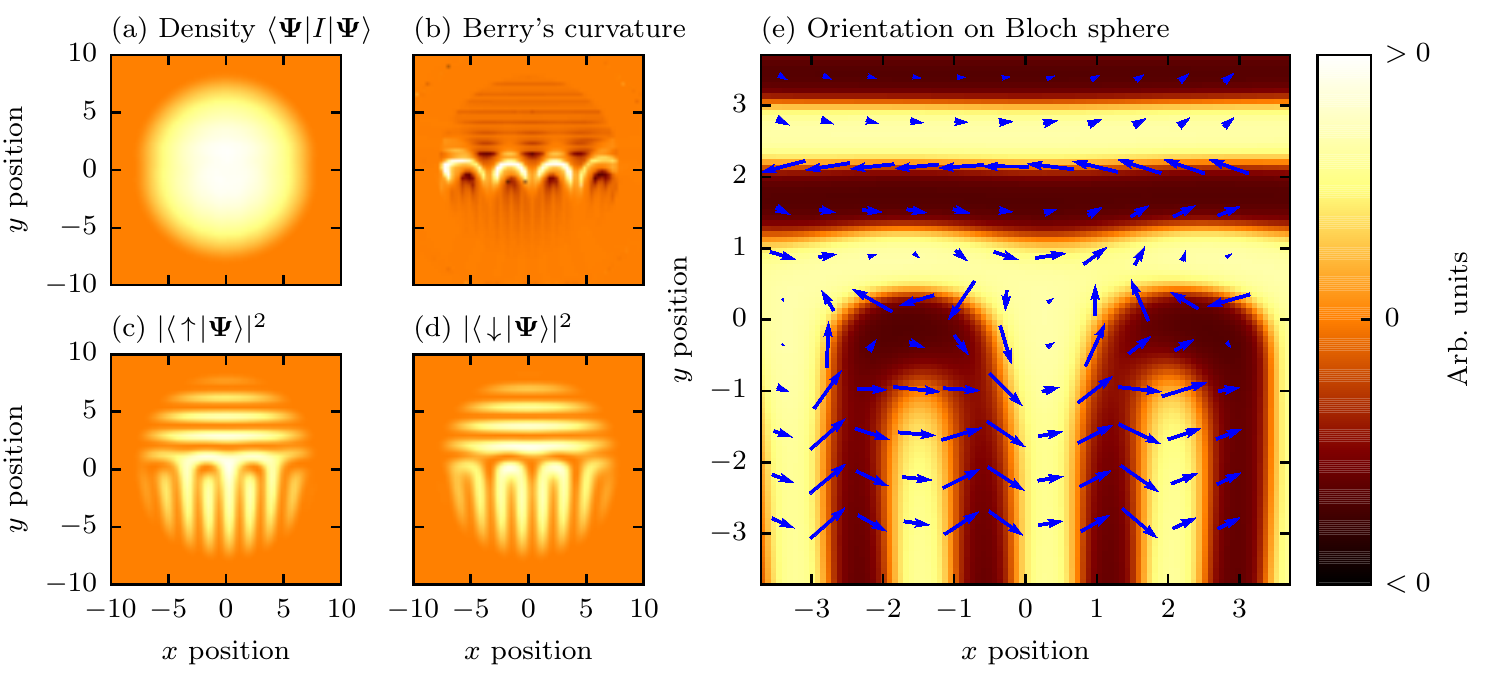}
\end{center}
\caption{The ground state of a spatially dependent SOC BEC with orthogonal
stripe patterns. This simulation was performed with $\mu=32\hbar\omega$,
$\tilde\delta^\prime=\Omega/\sqrt{2}$, $\kappa=2$ and $\theta=\pi/2$. In this
case, the boundary occurs at $x=0$. Panels (a)--(d) plot the density $\bra{\bPsi}I\ket{\bPsi}$, the Berry's curvature ${\mathcal B}_z$, and the densities in the spin up and spin down components
$|\left<\uparrow\!|\bPsi\right>|^2$, and $|\left<\downarrow\!|\bPsi\right>|^2$.
A row of vortices forms to link the two completely incompatible SW
patterns.  (e) Orientation of the local spin vector on the Bloch sphere.  The colored background gives the $\ez$ component, while the vector field plots the $\ex$ and $\ey$ components.}
\label{x} 
\end{figure}

Next we investigate how the ground-state density and phase profiles vary over
the interface region for the mismatched condition when two stripes in one domain
are linked to a single stripe in the other. Specifically, we consider
$\tan\theta=1/2$ and $\tilde\delta^\prime=\Omega\sqrt{10}/6$, which gives a
separatrix $2x+y=0$. Keeping the same values of $\mu$, $\kappa$ we used for
$\theta=\pi/4$ case, the ground state is shown in Fig.~\ref{2x+y}. We observe
that, near the trap center, every two adjacent stripes in the region of $2x+y>0$
connect to one single stripe in the region of $2x+y<0$. Unlike the previous case
with $\theta=\pi/4$, where the density and phase profiles both vary smoothly
across the interface, in the current case the phases profile does not vary
smoothly across the interface. In the spin-projections shown in
Fig.~\ref{2x+y}e we see that vortices form at the boundary. In this
case, two stripes must merge---and also form an excess vortex---before a single stripe
crosses the domain boundary. The existence of an imbalanced number vortices in the ground state
results from the merging of stripes, and is clearly visible in the Berry's curvature (Fig.~\ref{2x+y}b) which now favors negative values at the interface. In conventional BECs, vortices are
stabilized by the application of artificial magnetic fields. However in our
spatially dependent SOC BEC, the ground state supports vortices at the interface
between two distinct SW phases. Similar defect formation at the interface of two
distinct ground state phases was studied for spin-1 BECs with tunable inter- and
intra- species interactions \cite{Ruostekoski12PRL,Ruostekoski13PRA}.

Finally we examine the case where $\theta=\pi/2$ and $\tilde\delta^\prime=\Omega/\sqrt{2}$:
here the interface coincides with the $x$-axis. In this case the stripes for
$y<0$ are perpendicular to the interface while the stripes are parallel to the
interface in the region of $y>0$. The result is shown in Fig.~\ref{x}. The
orthogonal stripes result in the formation of a vortex chain on the interface
that can be seen clearly in the spin-projections of Fig.~\ref{x}e and in the Berry's curvature in Fig.~\ref{x}b. Our
results indicate that the unconventional BEC ground state contains chains of
vortices and anti-vortices stabilized by the position-dependent SOC.
Furthermore, the number of vortices is highly controllable by tuning the size of
condensate, the spin-orbit coupling strength $\kappa$, and the orientation of
the interface.

\section{Concluding remarks}

\label{sec:conclusion}

We proposed a new technique for creating position-dependent SOC for cold
atomic BECs.  This can be implemented by combining a cyclic Raman coupling scheme~\cite{Campbell2011} to
induce SOC, along with a magnetic field gradient~\cite{Spielman2009,Lin2009b} to impart a spatial dependance.

Subject to this combination, we find that a weakly interacting BEC phase separates
into two domains with orthogonally oriented stripes. Depending on axes of the
domain boundary---set by the spatial direction of the magnetic field
gradient---the stripes from each domain can intersect the boundary with
matched or mismatched spatial periods. We show that when the stripe
patterns intersect with different spatial periods, a chain of topological
defects, including vortices and anti-vortices, form to link the mismatched
stripe patterns. In contrast to vortices present in conventional rotating BECs,
here the vortices are stable topological defects that are not present in the
homogenous phase (here the SW phase). These vortices can form in an ordered
chain when the relative periods at the domain wall are different, but
commensurate, and they form a disordered chain when the relative periods are
incommensurate.

\ack

We appreciate enlightening conversations with our good friends. IBS
was partially supported by the ARO's atomtronics MURI, the AFOSR's
Quantum Matter MURI, NIST, and the NSF through the PFC at the JQI.
GJ acknowledges the support by the Research Council of Lithuania
(Grant No. MIP-082/2012). SCG and SWS are supported by the Ministry
of Sciences and Technologies, Taiwan (Grant No. MOST 103-2112- M-018-002-MY3
and 103-2923-M-007-001). LS would like to thank the support of the German Science 
Foundation (GRK 1729).

\section*{References}

\providecommand{\newblock}{}

\end{document}